# External quantum efficiency above 100% in photovoltaic cells due to the technique raising efficiencies for all kinds of solar cells

Jianming Li*

*Institute of Semiconductors, Chinese Academy of Sciences, A35 Qinghua East Road, Haidian District, Beijing 100083, P. R. China*

**ABSTRACT**

A V-shaped module (VSM) photovoltaic technique, which breaks traditional concepts, has been proven by European and American scientists to enable power-conversion-efficiency ($\eta$) to increase by 50% for thin-film solar cells. Furthermore, the VSM technique raises $\eta$ for all kinds of solar cells (J. Environ. Sci. Eng. A 12, 214 (2023). https://doi.org/10.17265/2162-5298/2023.06.002). Mysterious dark energy is thought to result in the surprising VSM effect. The VSM approach opens up new avenues to study photovoltaic physics. In this study, the external quantum efficiency (EQE) of commercial polycrystalline silicon solar cells in the VSM was investigated, which exhibits a surprising phenomenon of EQE above 100%. In theory, non-infrared light incident into a solar cell can cause infrared emission. The VSM could trap the emitted infrared photons and lead to extra photoexcited carriers. The easy-to-reproduce VSM effect is thus explainable. The energy of emitted infrared photons is considered as the so-called dark energy. This study provides new clues and evidences to unravel the mystery of the surprising VSM effect. The VSM technique could also be used to develop photodetectors for infrared and ultraviolet light respectively.

* Corresponding author.

*E-mail address:* jml_iscas@163.com (J. Li)； *Tel.:* +86-18310559676

## 1. Introduction

Today, the use of electric power is a basic need around the world. Since solar energy is plentiful and freely available, solar photovoltaic (PV) deployments provide a sustainable source of electricity [1]. In order to achieve optimum PV performance, it is essential to fully understand the physical principle of PV solar cells.

A V-shaped module (VSM) method [2], which breaks the traditional conceptions of PV technologies, has been proven to enable the power-conversion-efficiency ($\eta$) of polysilicon solar cells to raise from 13.4% to 20.2%, creating an $\eta$ increase of 51% [3]. Also, the VSM technique has been proven by European and American scientists to enable $\eta$ to increase by 50% for thin-film solar cells [4,5]. As the V-shaped structure makes use of light trapping and absorbs more photons, the VSM technique is effective for raising $\eta$ for all kinds of solar cells [6]. In addition, the research results of Canadian scientists have exhibited that the VSM arrays obtain substantially higher generated electrical power density than flat-panel arrays under direct sunlight irradiation [7].

The VSM approach has opened up new avenues for exploring the mysteries in photovoltaic physics. In this study, the VSM approach was used to further study the polycrystalline silicon solar cells through measuring external quantum efficiency (EQE) in detail.

## 2. Experimental procedures

For the VSM structure in this study, two solar cells with identical characteristics were placed at an angle to each other. The two cells were 1-cm-square commercially available polycrystalline solar cells with an $\eta$ of 13.4%, and these two cells were just the cells used in the work described in Ref. 3. Figure 1 shows the sketch of the VSM and light ray in the VSM, where the opening angle between the two tilted cells is $\alpha$.

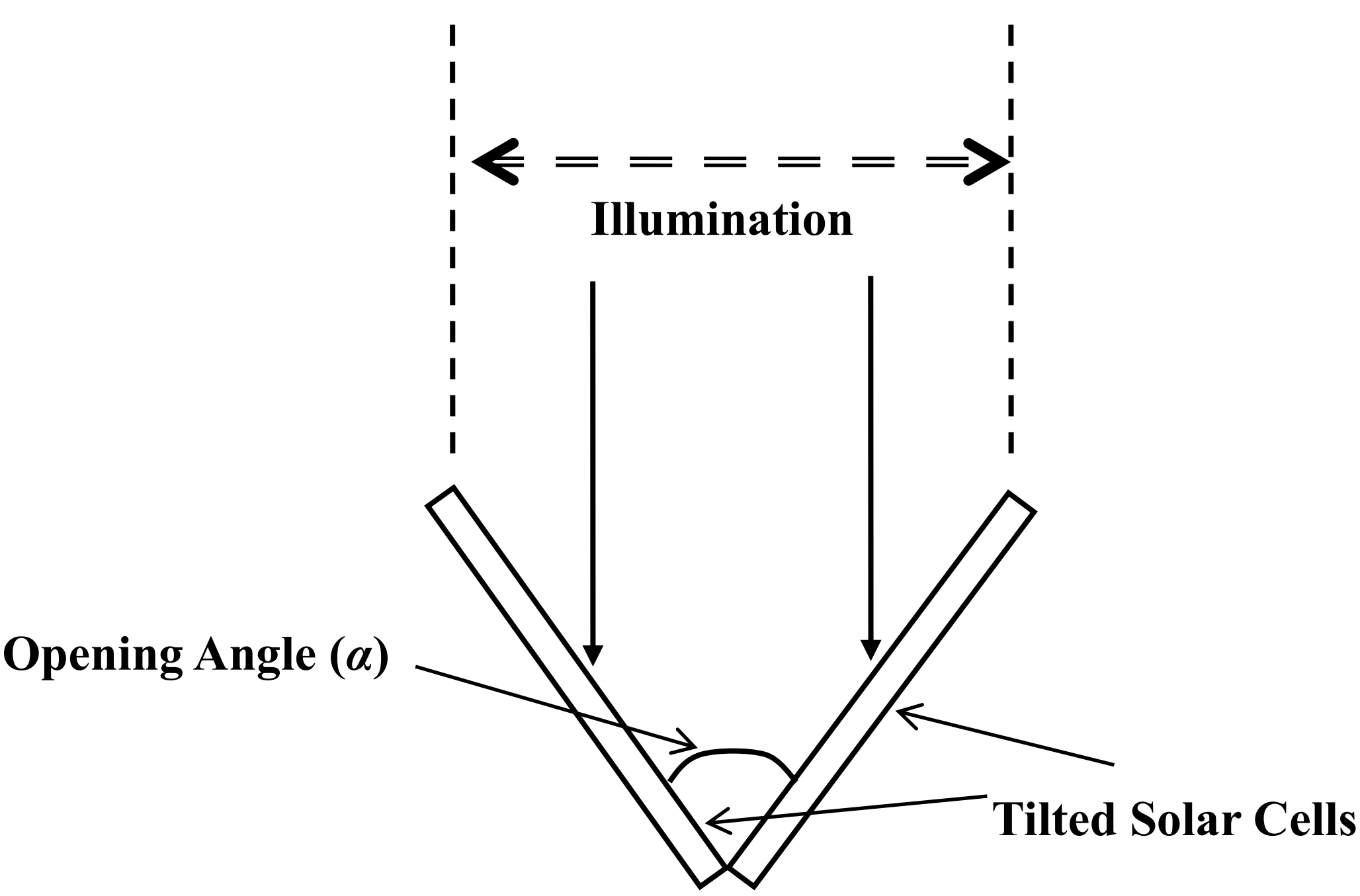

**Fig. 1.** A schematic cross section of the VSM structure.

In this work, EQE of the solar cells in the VSM was systematically investigated. EQE is defined as the ratio between the number of the photoexcited electrons flowing out of a solar cell and the number of illuminated photons upon it at a given wavelength, and EQE is thus incident photon conversion efficiency. The electrons flowing out of the cell is just the photoexcited carriers collected by an external circuit. For the tilted cells in the VSM, EQE is the ratio between the number of the collected carriers per projected illuminated area to the number of photons illuminated upon the cells per projected illuminated area.

In this study, EQE of the cells in the VSM was measured for four $\alpha$ values (20º, 30º, 40º, and 180º) respectively by using a spectral response measuring equipment (Model CEP-25/CH) made by Japan Bunkou Keiki Co., Ltd. The measurement error for the measuring equipment was below 2%.

## 3. Results and discussion

One of the cells in the VSM was measured first. Figure 2 shows EQE results. It is noted that the VSM approach enhanced EQE. The EQE values up to 1.8 (or 180%) and 1.4 (or 140%) were observed for $\alpha$ values of 20º and 30º, respectively.

The second cell in the V-shaped structure was measured for the same $\alpha$ as the first cell, and the two cells had consistent result. Furthermore, the two cells under parallel connection were measured in the same way as each cell. The performance of the two cells under parallel connection was consistent with that of each cell. It was very easy to reproduce the EQE results.

The EQE values above 100% appear very surprising in view of traditional solar cell theories. In order to get reliable data, a serious proceeding for the measurement was carried out [8].

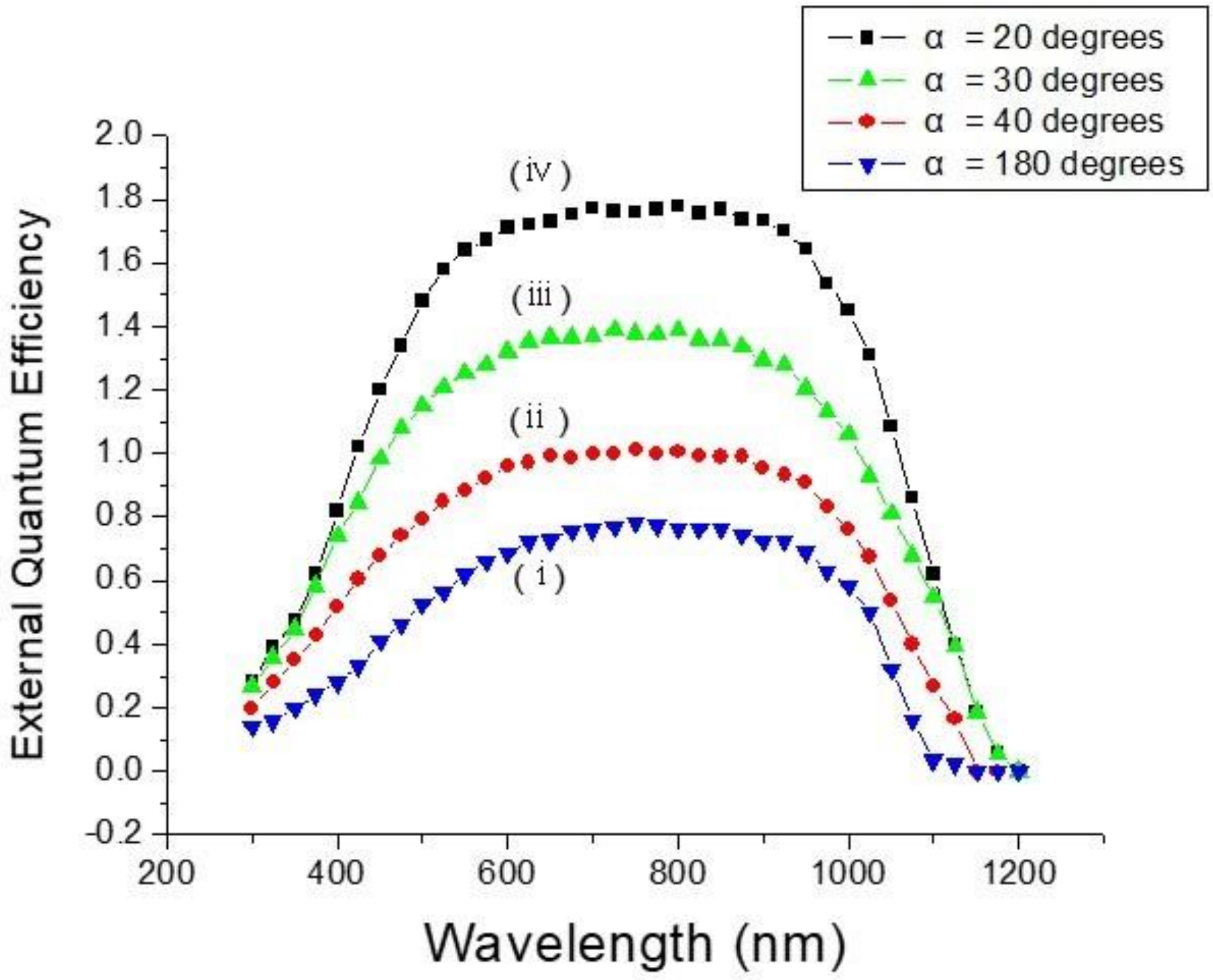

**Fig. 2.** The EQE results of the cells in the VSM.

Now, let us review the history of solar cell research. In principle, electron-hole pair multiplication (EHPM), the creation of more than one electron-hole pair from one absorbed photon, can occur in semiconductors when the absorbed photon has enough energy. Several leading research groups reported EQE values above 100% in quantum dot and organic solar cells respectively [9-13]. Their findings demonstrate that EHPM events occurred in these two kinds of cells. Furthermore, EQE above 100% in other solar cells is not impossible.

While investigating the EQE results in this work, the reflection from each cell in the VSM is considered first.

It is known that there is no antireflection coating (ARC) with zero reflectivity. In addition, the metal contacts of solar cells reflect incident light. The reflections from the cells result in optical loss, causing reduction in EQE.

For the VSM under illumination, the incident light is partially absorbed at one cell and the rest of light is reflected to the other cell, where the reflected light has another chance to be absorbed. Although each cell reflects some incident light, it receives the same amount of the reflected light from another cell that is also under illumination.

The density of the reflections in the VSM increases rapidly as $\alpha$ decreases, which gives rise to the light trapping effect. At small $\alpha$ values, the optical reflectance of the cells in the V-shaped geometry indicates an almost complete absorption of photons, and this structure is more reminiscent of the blackbody absorbers of early quantum physics.

It is well known that solar spectrum has an infrared (IR) component. The rear metal contacts of the solar cells in the VSM can serve as back-side mirrors for IR photons, enabling some IR photons of sunlight to bounce off the V-shaped cells many times. This geometry promotes IR absorption via optical path length enhancement within the cells, resulting in the better use of the solar spectrum.

In the V-shaped configuration, light is trapped due to the reflections from both cells as a result of which the absorption is enhanced. Thus, the VSM approach reduces reflection loss and leads to an enhancement in EQE compared to planar cells.

As described in Ref. 3, the reflectance of the cells used in this study is around 20%, resulting in about 20% optical loss. The EQE data achieved in this study show that the peak EQE of the cells in the VSM raises from 0.8 at $\alpha = 180^{\circ}$ (or planar cells) to 1.4 at $\alpha = 30^{\circ}$, increasing by 75%. For $\alpha = 20^{\circ}$, the peak EQE is even up to 1.8, increasing by 125% compared to planar cells. It can be seen that the VSM technique can significantly raise EQE values of the cells, creating EQE increments by exceeding 20%. Accordingly, the significant EQE increments cannot be solely attributed to the reduction of the reflection loss. It is believed that other factors also make contributions to the significant EQE increments.

From the EQE data, no EHPM occurred in the VSM with $\alpha = 180^{\circ}$ (planar cells). However, the EQE values above 100% demonstrate that EHPM events occurred in the VSM with $\alpha = 20^{\circ}$ and $\alpha = 30^{\circ}$ respectively. In order to explain the surprising EQE phenomena, some hypothetical factors are proposed as follows.

For the case in which an illuminated solar cell is in equilibrium, the incident energy falling from the sun upon the cell must equal the total energy output from the cell. Thus, a portion of photon energy falling upon the solar cell is converted into electrical output from the cell, and the rest of the energy must be released from the cells due to the constraint of energy conservation.

Now, let us consider the energy released from solar cells. In solar cells, the absorbed photons produce transitions between two sets of electronic states. When electrons are excited by photons with energy in excess of a forbidden energy band-gap ($Eg$), they tend to rapidly thermally relax to the conduction band edge, as shown in Fig. 3. The excess energy of carriers generated by absorption of photons with energy greater than $Eg$ is usually dissipated as heat, making the cells warmer. In addition, free carrier absorption can make the free carriers obtain energy, and the carriers in turn dissipate their excess energy to heat by interaction with the lattice. In a word, some of the incident energy into solar cells is lost as heat. It is well known that heat generates IR emission which is called as thermal radiation. This means that solar cells under illumination emit IR photons, leading to the release of energy.

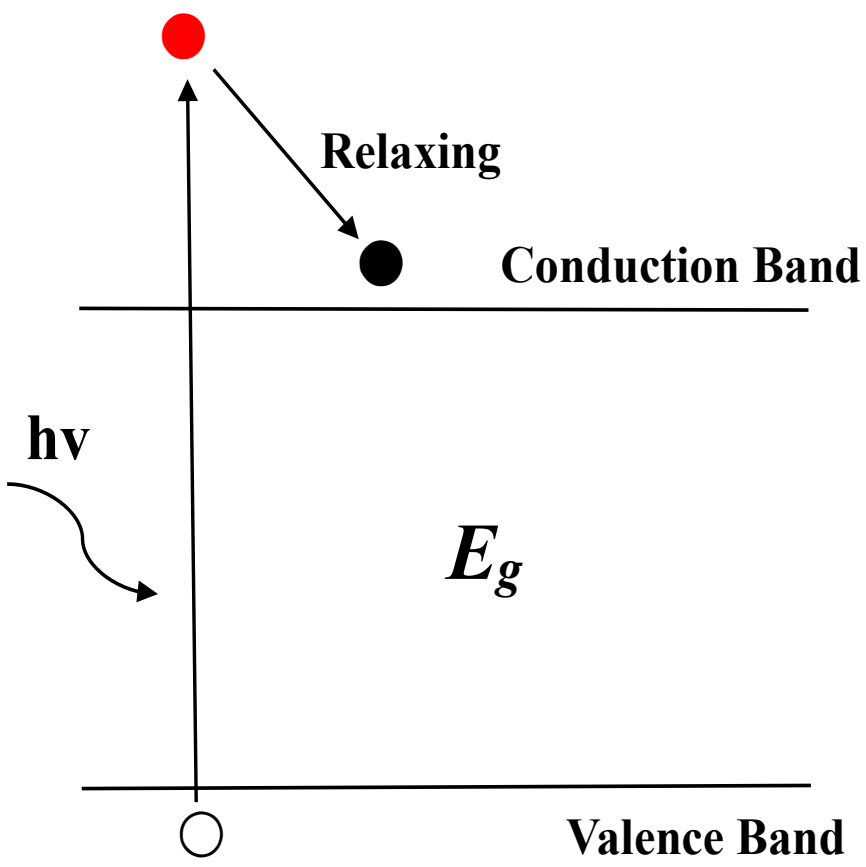

**Fig. 3.** A photoexcited electron relaxing to the conduction band edge.

As described above, the VSM traps the IR photons of sunlight due to optical confinement scheme. Similarly, some IR photons emitted from the cells can also bounce off the V-shaped cells.

Lattice imperfections lead to defect energy levels in *Eg*. The defect energy levels enable IR photons to be absorbed for creating carriers through sub-band-gap excitation. For example, the electron bound by an impurity (or a defect) can become a photoexcited electron, resulting in a positively charged defect left, as shown in Fig. 4.

The numerous defects in the grain boundaries of the polycrystalline silicon solar cells just create defect energy levels in *Eg*. For the VSM, non-IR light incident into each cell could cause IR emission, and the emitted IR photons could subsequently be absorbed via the defect energy levels to create extra carriers in the opposite cell. This means that EHPM theoretically can occur in the VSM. Therefore, the EQE values exceeding 100% achieved in this study are explainable.

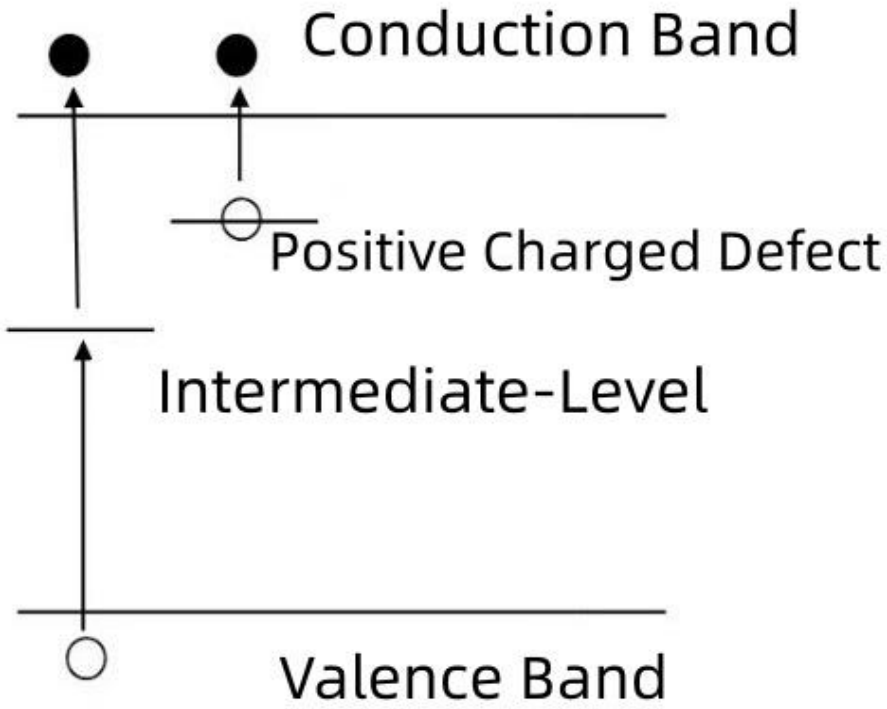

**Fig. 4.** The sub-band-gap excitation mechanism via the intermediate levels within forbidden gap.

For the VSM in this study, the optical trapping scheme makes both solar cells enhance each other equally, causing significant enhancement of EQE.

Furthermore, IR-photocurrent due to sub-band-gap excitation has been used to explain circuit devices, e.g. tunnel and Zener diodes as well as ohmic contacts [14]. Especially, the IR-photocurrent explanation breaks the paradigm of quantum mechanics dominating the interpretation of ohmic contact as well as Zener and tunneling diodes, indicating the wave-particle duality of electrons in semiconductor circuit devices. The results of this study just support the viewpoint that IR-photocurrent is created due to IR absorption through sub-band-gap excitation in semiconductor devices.

## 4. Conclusion

In conclusion, the IR emission of solar cells is believed to be one of the major loss mechanisms by which photon energy is wasted in planar modules. However, the VSM technique can substantially reduce these energy losses through trapping the IR emission. The energy of the emitted IR photons is dark energy that could lead to significant increases in both EQE and $\eta$ of the solar cells in the VSM. The EQE values above 100% achieved in this study are the evidences supporting the hypothesis that non-IR light incident into each cell in the VSM leads to IR emission which creates extra carriers in the facing cell. According to this study, the surprising VSM effect is mainly attributed to the hypothetical factors described above.

## Declaration of competing interest

This is to certify that this manuscript does not make any conflict of interest with any person or institution or laboratories or any work.

## Data availability

The data that support the findings of this study are available from the corresponding author upon reasonable request.

## Acknowledgements

This work was financially supported by Chinese Academy of Sciences (CAS). The author is grateful to Mr. Haitao Liu and Mr. Yonghui Zhai of the Photovoltaic and Wind Power Systems Quality Test Center, CAS for their efforts in the EQE measurements.

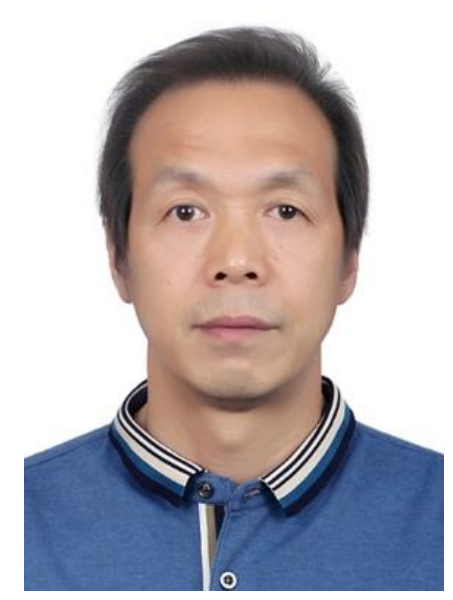

Jianming Li, research associate and senior engineer at the Institute of Semiconductors, Chinese Academy of Sciences. Email: jml_iscas@163.com; Tel.: +86-18310559676

Jianming Li was born in Beijing, China, in 1961. He studied at Beijing Normal University from 1980 to 1987. After graduated from the University, he joined the Institute of Semiconductors, Chinese Academy of Sciences. Since 1987, he has been engaged in research on photovoltaics, semiconductor materials and devices, etc. As a Visiting Scientist, he worked at Brookhaven National Laboratory, USA, from 1992 to 1994. As a Visiting Scientist, he worked at Pennsylvania State University, USA in 1997.